\documentclass[submitting, onecolumn]{nst}

\usepackage{subfigure,dcolumn}
\usepackage[T2A,T1]{fontenc}
\usepackage[russian,english]{babel}
\usepackage{multirow}
\usepackage{makecell}

\usepackage{listings}
\begin{document}

\title{Radiation hardness study of BC408 plastic scintillator under 80 MeV proton beam irradiations}

\author{Yue Zhang}
\affiliation{Institute of High Energy Physics, Chinese Academy of Sciences, Beijing 100049, China}
\affiliation{China Spallation Neutron Source, Dongguan 523803, China}
\affiliation{Shanghai Advanced Research Institute, Chinese Academy of Sciences, Shanghai 201210, China}

\author{Ruirui Fan}
\email[Corresponding author, ]{Ruirui Fan, fanrr@ihep.ac.cn, 1 Zhongziyuan Road, Dalang, Dongguan ,Guangdong 523000, China}
\affiliation{Institute of High Energy Physics, Chinese Academy of Sciences, Beijing 100049, China}
\affiliation{China Spallation Neutron Source, Dongguan 523803, China}
\affiliation{State Key Laboratory of Particle Detection and Electronics, Beijing, 100049, China}

\author{Yuhong Yu}
\email[Corresponding author, ]{Yuhong Yu, yuyuhong@impcas.ac.cn, 509 Nanchang Road, Lanzhou, China}
\affiliation{Institute of Modern Physics, Chinese of Academy of Sciences, Lanzhou 730000, China}

\author{Hantao Jing}
\affiliation{Institute of High Energy Physics, Chinese Academy of Sciences, Beijing 100049, China}
\affiliation{China Spallation Neutron Source, Dongguan 523803, China}

\author{Zhixin Tan}
\affiliation{Institute of High Energy Physics, Chinese Academy of Sciences, Beijing 100049, China}
\affiliation{China Spallation Neutron Source, Dongguan 523803, China}

\author{Yuhang Guo}
\affiliation{Institute of High Energy Physics, Chinese Academy of Sciences, Beijing 100049, China}
\affiliation{China Spallation Neutron Source, Dongguan 523803, China}

\author{You Lv}
\affiliation{Institute of High Energy Physics, Chinese Academy of Sciences, Beijing 100049, China}
\affiliation{China Spallation Neutron Source, Dongguan 523803, China}

\begin{abstract}

To investigate the 1.6 GeV high-energy proton beam detector utilized in the CSNS Phase-II upgrade project, a plastic scintillator detector presents a viable option due to its superior radiation hardness. 
This study investigates the effects of irradiation damage on a BC408 plastic scintillator induced by 80 MeV protons, including absorption and fluorescence spectroscopy, and light yield tests of BC408 pre- and post-proton irradiation, with a focus on determining the radiation resistance threshold of BC408. The results indicate that the performance of BC408 remains unimpaired at absorbed doses up to $5.14\times10^3$ $\mathrm{Gy/cm^3}$, demonstrating its ability to absorb $1.63\times 10^{13}\,\mathrm{p/cm^3}$ 1.6 GeV protons while maintaining stability. This suggests that BC408 could potentially be used as the 1.6 GeV high-energy proton beam detector in the CSNS Phase-II upgrade project.

\end{abstract}
\maketitle


\section{Introduction}\label{sec:introdution}

Plastic scintillator is an organic material that produces a luminescence effect when it interacts with ionizing radiation\cite{kharzheevRadiationHardnessScintillation2019a,salimgareevaPlasticScintillatorsBased2005}. Incident radiation causes $\pi$-electron excitation in scintillator materials and then relaxes to their ground state through the fluorescence process of light emission. The amount of light emitted depends on the energy of the excited particles, so the scintillator can be used for particle identification. In addition, incident radiation will lead to the formation of ions and free radicals in the process of energy dissipation, which affect the luminous ability of scintillators and damage the molecular structure\cite{brossRadiationDamagePlastic1992,zornPedestrianGuideRadiation1993,mizuehamadaRadiationDamageStudies1999}.

The plastic scintillator consists of organic flours suspended in the polymer base, which contains some form of aromatic ring structure that contains $\pi$-electron. Organic flours consist of primary flours and secondary flours. The primary flour makes the plastic scintillator transparent to its own scintillation light reducing self-absorption and the secondary flours act as a wavelength shifter\cite{kharzheevRadiationHardnessScintillation2019a}. This wavelength of light is more transparent to scintillators and has a better response to the photocathode of photomultiplier tubes (PMT), which improves light collection efficiency.

\begin{figure}[!htbp]
  \centering
  \includegraphics[width=0.6\textwidth]{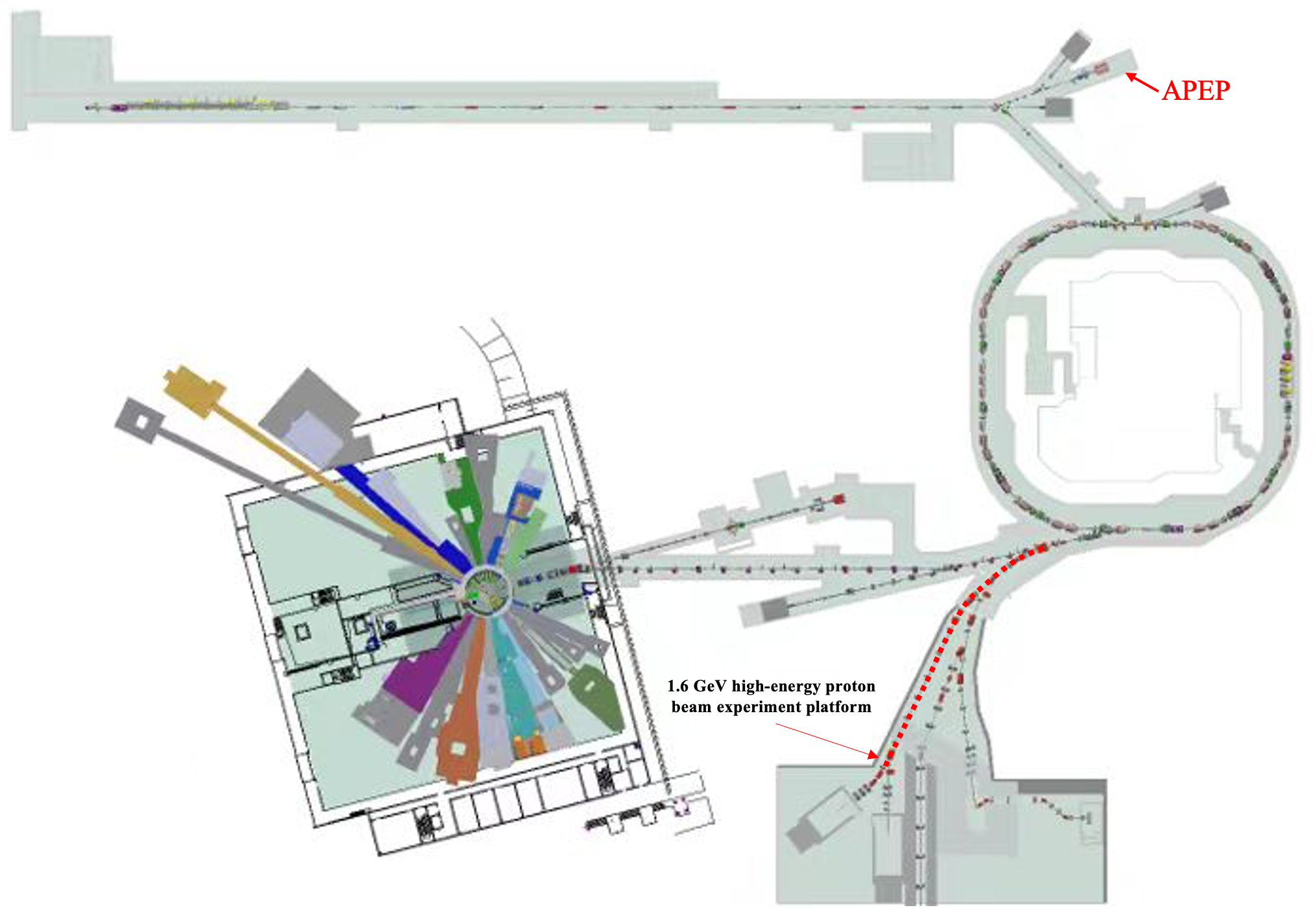}
  \caption{The 1.6 GeV high-energy proton beam experiment platform in CSNS-II-Up.}
  \label{Fig1_RPP}
\end{figure}

Plastic scintillator is an excellent material for measuring charged particles such as $\alpha$, $\beta$, protons and fission fragments with 100\% detection efficiency and plays an important role in the field of detector physics\cite{kharzheevRadiationHardnessScintillation2019a,beddarPlasticScintillationDosimetry2006,dongChargeMeasurementCosmic2019}. Because the plastic scintillator is mostly composed of low atomic number materials, it has low detection efficiency for $\gamma$ rays. These characteristics are usually used to measure charged particles under strong $\gamma$ conditions and a good repulsion ratio can be obtained. The plastic scintillator has better radiation hardness than others, and a short decay time, so it is mainly used for intensity detection and Time-of-Flight (TOF) counters\cite{linPlasticScintillationDetectors2017}. The China Spallation Neutron Source Phase-II upgraded (CSNS-II-Up) plans to build a 1.6 GeV high-energy proton beam experiment platform with a design current of $10^8\,\mathrm{p/cm^2/s}$ (Figure \ref{Fig1_RPP}), so it is necessary to design a high energy proton beam intensity detector. As mentioned above, plastic scintillator is a potential candidate material. At the same time, considering that the high-energy proton beam may damage the molecular structure of the scintillator and the damage results in a significant decrease in the light yield of the scintillator and the introduction of errors in the captured data, it is necessary to study the effect of proton irradiation on the plastic scintillator to help the design and application of the subsequent high-energy proton beam detector.

In this study, 80 MeV proton beam was used to study the radiation hardness of plastic scintillators. Light yield testing in response to $^{241}\mathrm{Am}$, absorption spectroscopy and fluorescence spectroscopy were used to evaluate the performance changes of the plastic scintillator before and after proton irradiation.

\section{Experimental setup}
\subsection{The Associated Proton beam Experiment Platform}
According to LISE++ calculation, the deposition energy of 10 mm thickness of plastic scintillator irradiated by 1.6 GeV proton is 2.02 MeV and the deposition energy of 80 MeV proton is 9.09 MeV\cite{tarasovLISERadioactiveBeam2008}. The absorbed dose is the average radiation energy absorbed per unit mass of irradiated material, which is suitable for comparing different types of radiation and irradiated substances. Base on the above calculations, the energy deposited by a 80 MeV proton in a 10 mm thickness plastic scintillator is equivalent to 4.5 protons of 1.6 GeV.

\begin{figure}[!htbp]
  \centering
  \includegraphics[width=0.6\textwidth]{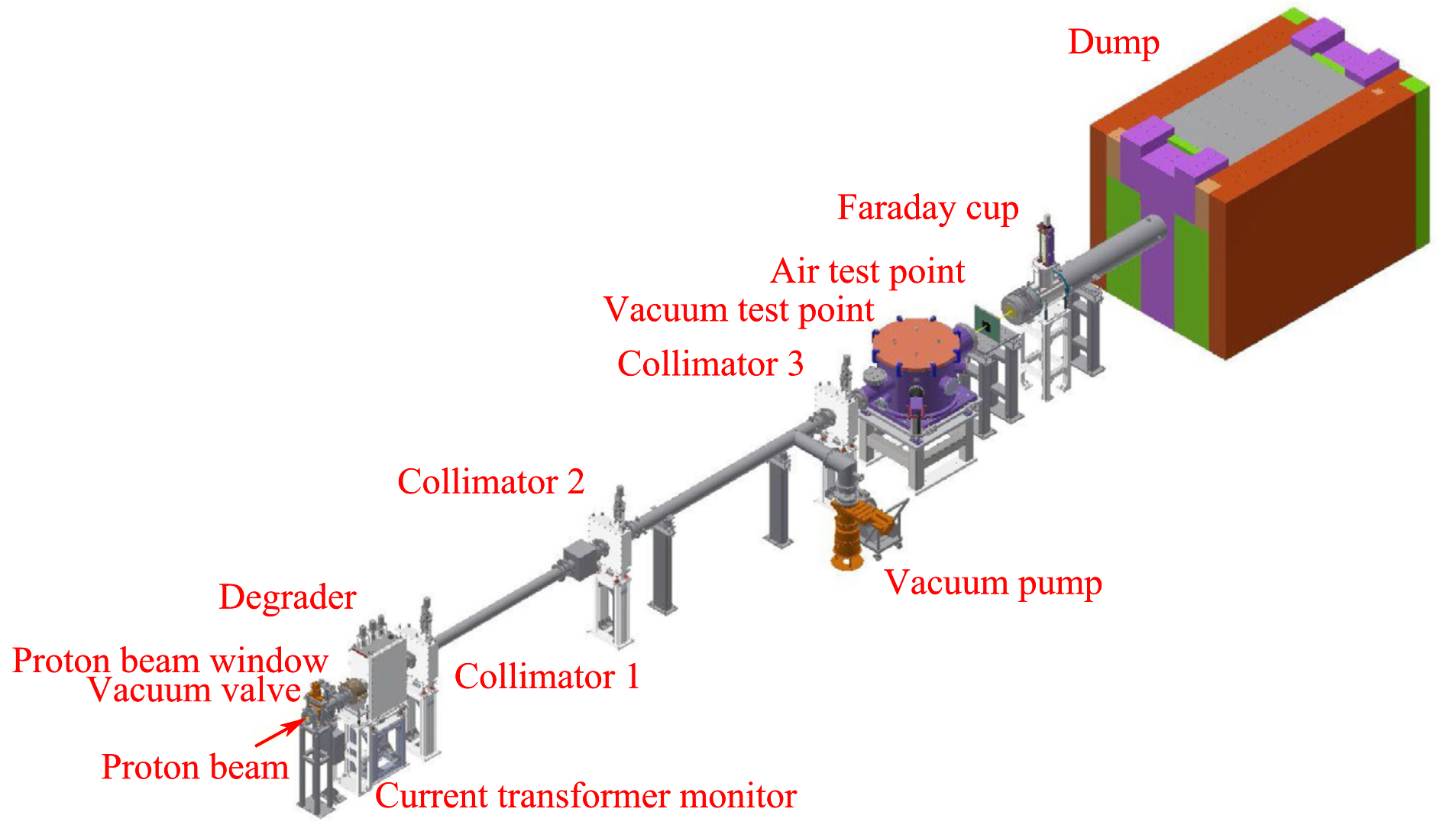}
  \caption{The main composition of the Associated Proton beam Experiment Platform\cite{liuPhysicalDesignAPEP2022a}.}
  \label{Fig3_proton_source_en}
\end{figure}

In this study, high flux 80 MeV proton beams were used to irradiate plastic scintillators. 
As shown in Figure \ref{Fig3_proton_source_en}, the Associated Proton Beam Experimental Platform (APEP) beam line is located at the end of the CSNS linear accelerator (Figure \ref{Fig1_RPP})\cite{liuPhysicalDesignAPEP2022a}. It is used to carry out experiments and test studies related to the application of proton irradiation.
Figure \ref{Fig3_proton_source_en} is a schematic diagram of APEP, which can provide proton beam energy from 10 to 80 MeV with the square beam spot size of $10\times10\,\mathrm{mm}^2$ to $50\times50\,\mathrm{mm}^2$. The proton flux is $10^7$-$10^{10}\,\mathrm{p/cm^2/s}$. When the proton energy is 80 MeV, the proton flux is the maximum. When CSNS power is 140 kW, the 80 MeV proton flux is $1.2\times10^{10}\,\mathrm{p/cm^2/s}$. APEP has two test points, a vacuum test point and an air test point. This study was carried out at the air test point\cite{liuPhysicalDesignAPEP2022a}.

\subsection{Sample preparation}
\begin{figure}[!htbp]
  \centering
  \includegraphics[width=0.6\textwidth]{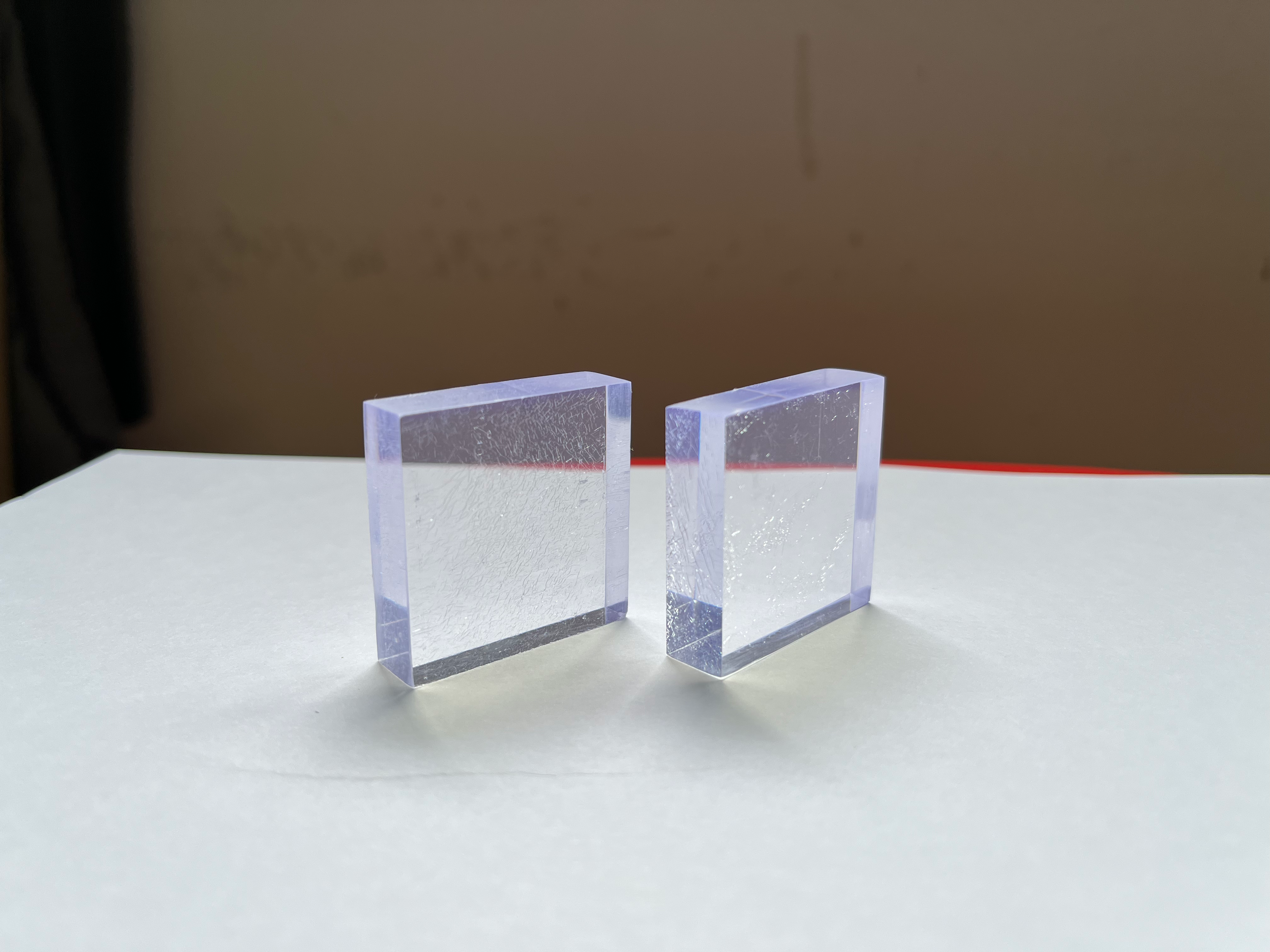}
  \caption{The color of sample BC408 under sunlight.}
  \label{Fig2_BC408_2}
\end{figure}

\begin{figure}[!htbp]
  \centering
  \includegraphics[width=0.6\textwidth]{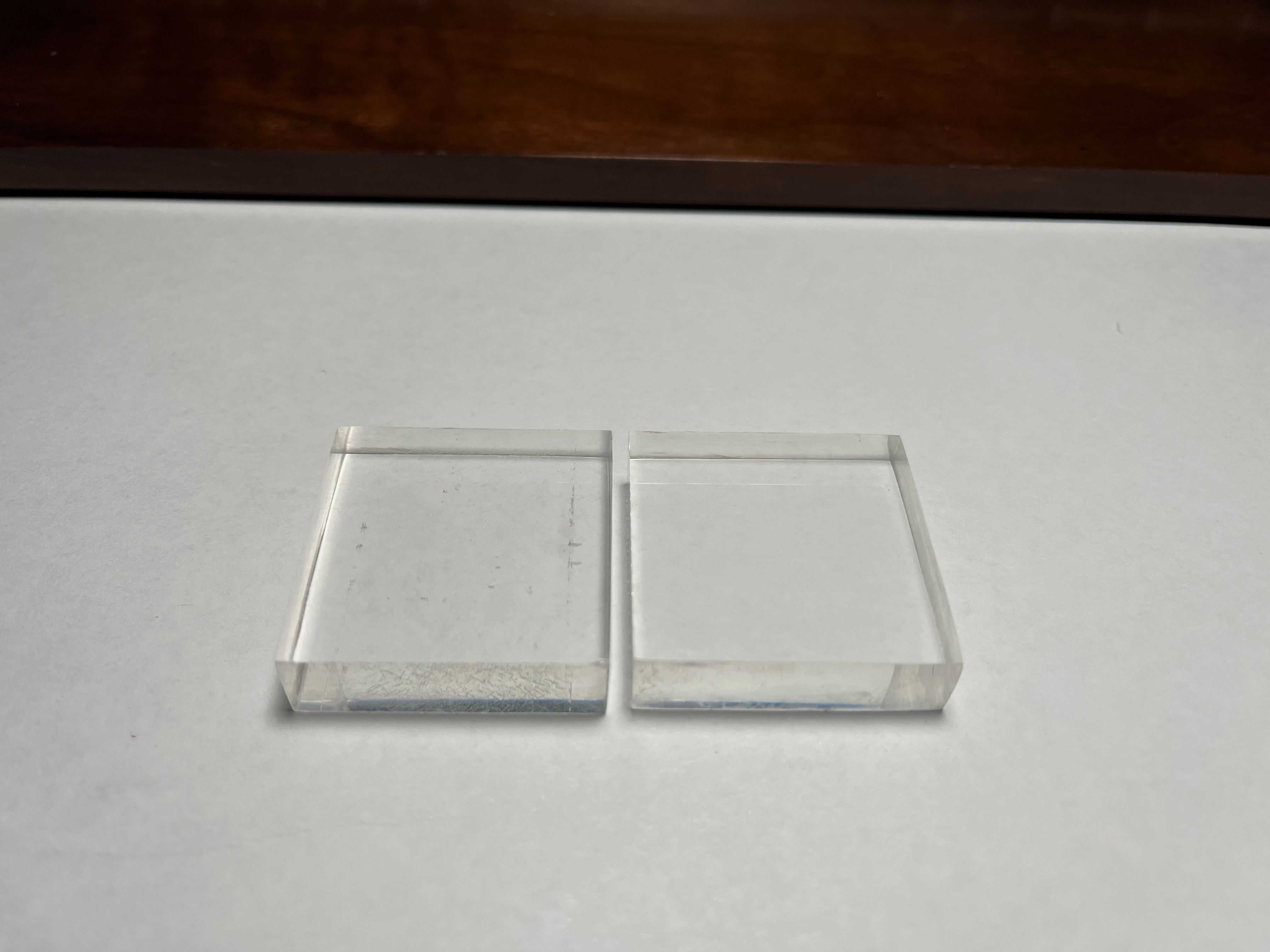}
  \caption{The color of sample BC408 under a fluorescent lamp.}
  \label{Fig2_BC408_1}
\end{figure}

The plastic scintillators used in this study were BC408, which is manufactured by Saint Gobain Crystals and their properties are outlined in Table \ref{tab:BC408_properties}. BC408 is a commercially-available plastic scintillator based on polyvinyltoluene, with high light output, good light transmission and short decay time, and is economical for use in large areas. In this study, eight BC408 samples with a size of $45\,\mathrm{cm} \times 45\,\mathrm{cm} \times10\,\mathrm{cm}$ were used, of which six pieces, named Plastic1-6, were irradiated by protons for different times as the experimental group and the other two pieces, named Plastic9-10, no irradiation as control groups that can be used as a reference for changes in BC408 quality and performance before and after irradiation. eight BC408 samples were stored in the same environment, except that Plastic1-6 scintillators were irradiated by protons, to minimize the influence of inconsistent environmental background. The color of the BC408 sample is transparent lavender under sunlight (Figure \ref{Fig2_BC408_2}) and colorless transparent under fluorescent light (Figure \ref{Fig2_BC408_1}).

\begin{table}[!ht]
  \centering
  \caption{The properties of BC408 scintillators}
  \begin{tabular}{l|l|l|l|l|l}
    \toprule
    Manufactured          & Base & \makecell{Light Output \\ (\% Anthracene)} & \makecell{Wavelength of Max.\\ (Emission, nm)} & \makecell{Rise Time \\(ns)} & \makecell{Decay Time \\(ns)}\\ \midrule
    Saint Gobain Crystals & PVT  & 64 & 425 & 0.9 & 2.1 \\ \bottomrule
  \end{tabular}
  \label{tab:BC408_properties}
\end{table}

In table \ref{tab3:PlasticSetup}, the 80 MeV proton beam with the maximum flux was used in the experiment, and $50\times50\,\mathrm{mm}^2$ square beam spot is set to cover the entire incident surface of the plastic scintillators. The experimental group, six BC408, were irradiated for a minimum of 5 minutes and a maximum of 20 hours, respectively. Plastic6 irradiated in shortest 5 minutes is equivalent to 1.6 GeV proton irradiation in the order of magnitude $1.63\times10^{13}\,\mathrm{p/cm^2}$ and Plastic1 irradiated in longest 20 hours is equivalent to 1.6 GeV proton irradiation in the order of magnitude $3.90\times10^{15} \,\mathrm{p/cm^2}$. Other irradiation time settings are used to observe changes of plastic scintillators radiation damage.

 \begin{table}[!htbp]
  \centering
  \caption{Comparison of irradiation time and absorbed dose of BC408}
  \label{tab3:PlasticSetup}
  \begin{tabular}{l|l|l|l|l|l|l}
  \toprule
  \multirow{2}{*}{\textbf{Plastic ID}} & 
  \multirow{2}{*}{\makecell{\textbf{Thickness}\\ \textbf{(mm)}}} &
  \multicolumn{3}{c|}{\textbf{80 MeV Proton}} & 
  \multicolumn{2}{c}{\textbf{1.6 GeV Proton}} \\ \cline{3-7}
  & &
  \makecell{\textbf{Irradiated} \\ \textbf{Time}} &
  \makecell{\textbf{Energy Deposition}\\ \textbf{(MeV/p)}} &
  \makecell{\textbf{Absorbed Dose}\\($\mathrm{\mathbf{Gy/cm^3}}$)} &
  \makecell{\textbf{Energy Deposition}\\\textbf{(MeV/p)}} &
  \makecell{\textbf{Equal 1.6 GeV}\\\textbf{Proton counts ($\mathrm{\mathbf{p/cm^3}}$)}}\\ 
  \midrule
  Plastic10 &\multirow{8}{*}{10} & \multirow{2}{*}{0 mins} & \multirow{2}{*}{0} & \multirow{2}{*}{0} & \multirow{2}{*}{0} & \multirow{2}{*}{0} \\
  Plastic9  & & & & & & \\
  \cline{3-7}
  Plastic6 &                      & 5 mins  & \multirow{6}{*}{9.09} & $5.14\times10^3$& \multirow{6}{*}{2.02} & $1.63\times10^{13}$\\
  Plastic5 &                      & 24 mins &                       & $2.47\times10^4$&                       & $7.80\times10^{13}$\\
  Plastic4 &                      & 48 mins &                       & $4.94\times10^4$&                       & $1.56\times10^{14}$\\
  Plastic3 &                      & 4 hours &                       & $2.47\times10^5$&                       & $7.80\times10^{14}$\\
  Plastic2 &                      & 8 hours &                       & $4.94\times10^5$&                       & $1.56\times10^{15}$\\
  Plastic1 &                      & 20 hours&                       & $1.23\times10^6$&                       & $3.90\times10^{15}$\\ 
  \bottomrule
  \end{tabular}%
  \end{table}

\subsection{Absorption spectroscopy and fluorescence spectroscopy}

Ultraviolet-visible absorption spectroscopy (UV-Vis) and photoluminescence spectroscopy are the most basic techniques for characterizing an analyte\cite{shardChapterUltravioletVisible2020}. Both techniques were used to study the effects of radiation damage on the optical properties of samples. UV-vis absorption spectroscopy was determined by HATACHI UH4150. Light absorption spectroscopy was measured relative to absorption in air over a range of 200 nm to 800 nm. Fluorescence spectroscopy was determined by HATACHI F-7100. The instrument measuring parameters were excitation wavelength 300.0 nm, excitation slit 2.5 nm and emission slit 2.5 nm.

The spectrum analysis of plastic scintillators before and after proton irradiation is carried out, and when testing the absorption spectrum, by comparing the absorption spectrum of samples in different periods, it can be used to analyze the difference in light absorption of plastic scintillators and explain the change in the performance of plastic scintillators. When testing the fluorescence spectrum, it is necessary to compare the difference in fluorescence intensity in addition to the difference of peak position of the fluorescence spectrum before and after irradiation. 

Since fluorescence spectroscopy measurement is related to the lifetime of the instrument's emitting and receiving devices, fluorescence spectroscopy measured in different periods will be different because of the different lifetime of the instrument devices. Therefore, it would be better to compare the fluorescence spectroscopy in the same period. Because BC408 is a commercial plastic scintillator and has the same quality in the same batch, two unirradiated samples, Plastic9 and Plastic10, can be used as the control group, to compare fluorescence spectroscopy of irradiated samples, to show the effects of radiation damage on the optical properties of the samples.

\subsection{Light yield testing in response to $^{241}\mathbf{Am}$}

\begin{figure}[!htbp]
  \centering
  \includegraphics[width=0.6\textwidth]{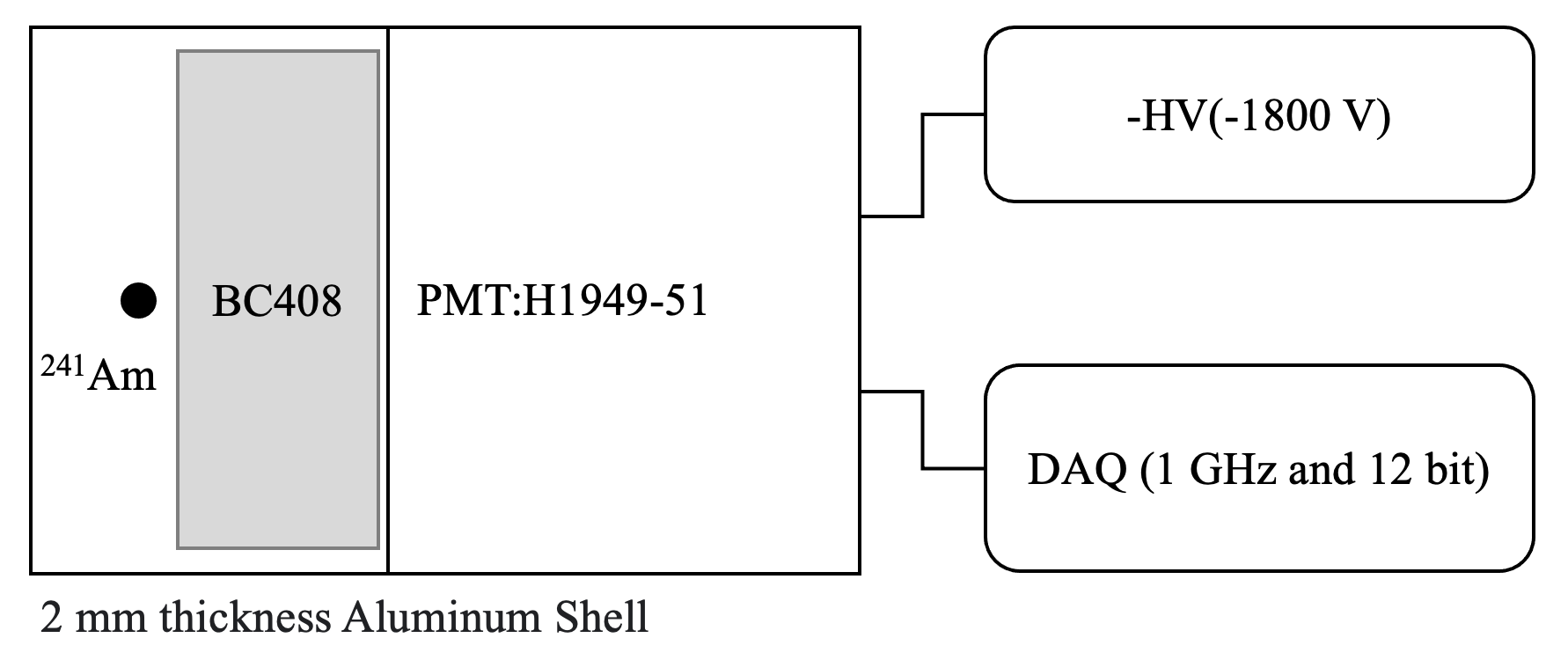}
  \caption{Light yield testing in response to $^{241}\mathrm{Am}$.}
  \label{Fig3_LY_layout}
\end{figure}

The light yield of the scintillator in response to $\alpha$ particles emitted by a $^{241}\mathrm{Am}$ source was measured using the device shown in figure \ref{Fig3_LY_layout}, which consists of an Aluminum shell, a standard $^{241}\mathrm{Am}$ source, a PMT, a high voltage power supply and a data acquisition system (DAQ).
The 2 mm thickness of the aluminum shell is used to avoid light. The PMT model is a H1949-51 PMT module produced by Hamamatsu Company, built-in PMT model R1828-01, its effective diameter is 46 mm, and its photocathode material is Bialkali. Its quantum efficiency curve matches the fluorescence spectrum of BC408 scintillator and its voltage divider design scheme facilitates electron collection, reduces transit time fluctuations, achives better time resolution and better output current linearity. DAQ is a waveform digital data acquisition system developed by the University of Science and Technology of China. This system is based on the PXIe system and has a sampling rate of 1 GHz, a 12 bit resolution, a transfer rate of 400 Mbps and a storage rate of 200 Mbps\cite{zhangSystemDesignPrecise2017,yuElectronicsTimeofFlightMeasurement2019}. 

The $\alpha$ energy spectrum of $^{241}\mathrm{Am}$ source is composed of 84.8$\%$ 5.5 MeV and 13.1$\%$ 5.4 MeV.
Considering the low energy resolution of the plastic scintillator, the position of the main energy peak of 5.5 MeV is taken as the reference point. In order to avoid the influence of silicone oil coupling, air coupling is adopted between plastic scintillators and PMT. The PMT output signal is directly input to the DAQ system recorded by the computer.

The experimental scheme of the light yield test is as follows: before proton irradiation, the response of eight samples to $^ {241}\mathrm{Am} $radiation source was tested, and the peak position of 5.5 MeV $\alpha$ particles was measured. After proton irradiation, the same test was performed, and the relative change in light yield was explained by comparing the peak position of 5.5 MeV $\alpha$ particles before and after irradiation.

\section{Results and analysis}
\subsection{Comparing sample irradiatied by proton}

\begin{figure}[!htbp]
  \centering
  \includegraphics[width=0.6\textwidth]{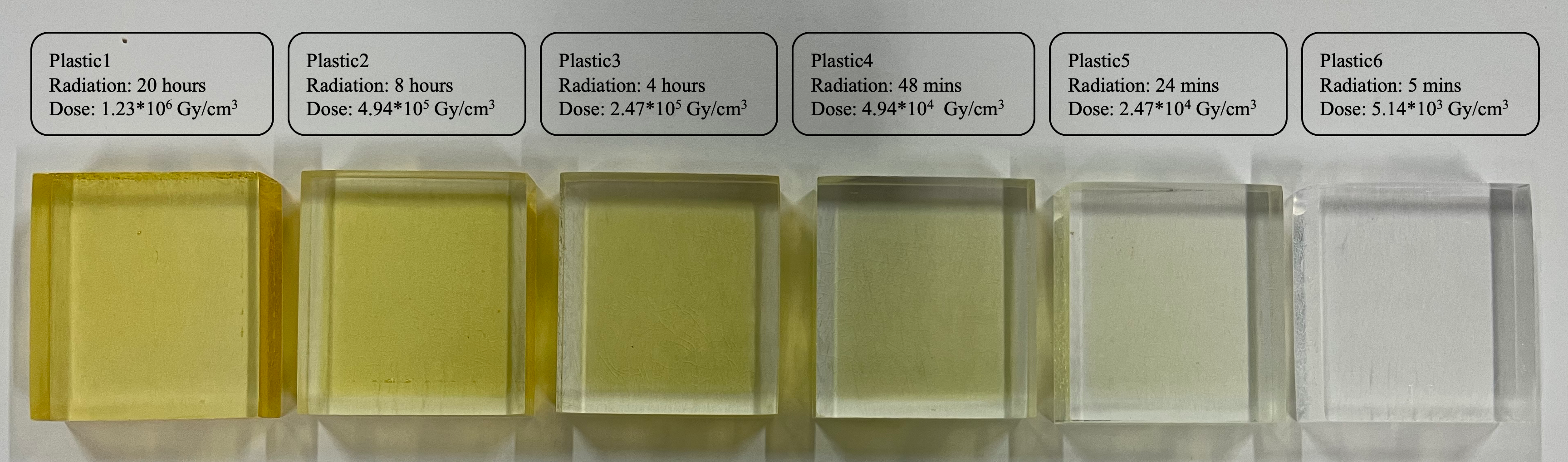}
  \caption{The color change of BC408 samples irradiated by proton.}
  \label{Fig3_BC408_color}
\end{figure}

\begin{figure}[!htbp]
  \centering
  \includegraphics[width=0.6\textwidth]{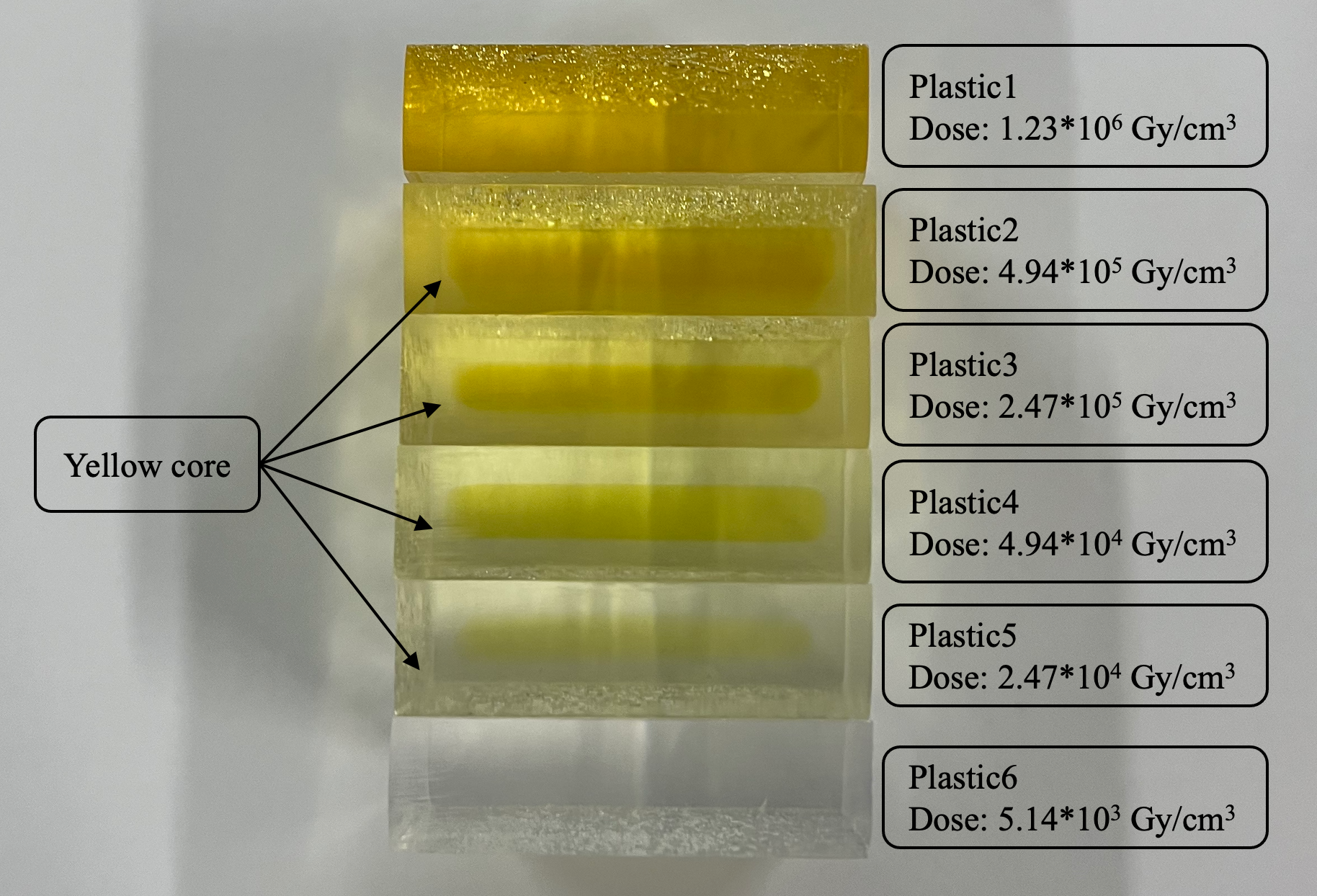}
  \caption{Appearing a color core after proton irradiation.}
  \label{Fig3_BC408_color_1}
\end{figure}

The color change of six BC408 samples irradiated by different doses of proton is shown in Figure \ref{Fig3_BC408_color}. The photo was taken under a fluorescent lamp. Obviously, as the proton irradiation dose increases, the color of the sample changes from colorless transparent to yellow. The higher the radiation dose, the darker the color of the sample. Compared to unirradiated samples (figure \ref{Fig2_BC408_1}), the sample is still colorless and transparent by $5.14\times10^3$ $\mathrm{Gy/cm^3}$ dose proton. Under the proton irradiation of $2.47\times10^4$ $\mathrm{Gy/cm^3}$ dose, the light yellow core appeared (figure \ref{Fig3_BC408_color_1}), and the color around the core changed from colorless transparent to light yellow with the increase of the radiation dose. When the sample is irradiated with $1.23\times10^6$ $\mathrm{Gy/cm^3}$ dose proton, the whole sample becomes uniform and dark yellow. 

\subsection{Absorption spectroscopy}

\begin{figure}[!htbp]
  \centering
  \includegraphics[width=0.6\textwidth]{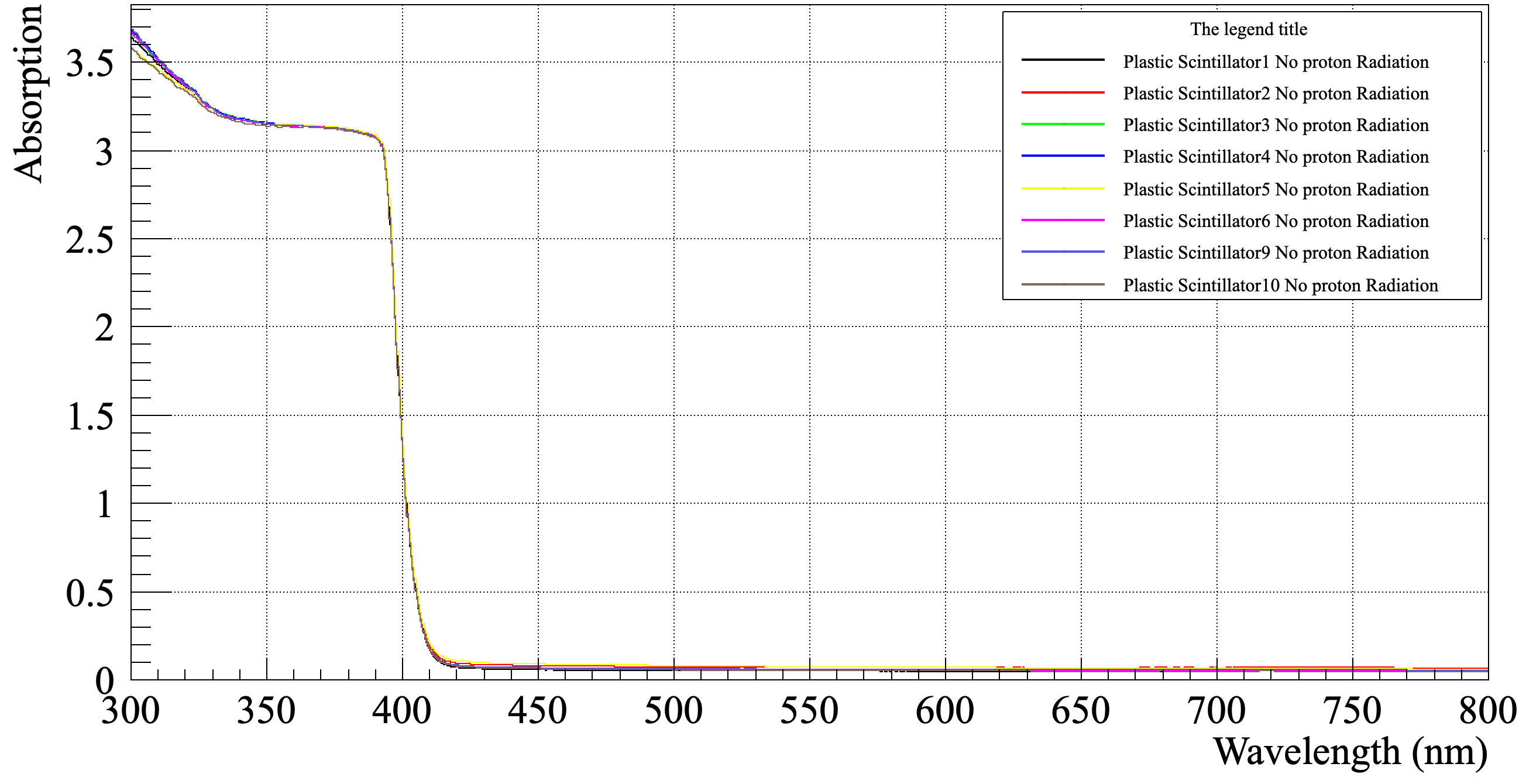}
  \caption{The absorption spectra of unirradiated BC408 samples.}
  \label{Fig3_Absor_Be}
\end{figure}

Figure \ref{Fig3_Absor_Be} illustrates the absorption spectra of eight BC408 samples post proton irradiation in the 200 nm to 800 nm wavelength range. The absorption spectra of the eight samples were consistent before proton irradiation. Within the 330 nm to 390 nm range, the absorbance value exceeds 3.1 and exhibits minor change, representing the shoulder peak of the absorption curve. Between 390 nm and 420 nm, the absorbance decreases rapidly from 3.1 to about 0.1. In the visible region beyond 420 nm, the absorbance is less than 0.1, signifying the sample's incapacity to absorb wavelengths within this range.

\begin{figure}[!htbp]
  \centering
  \includegraphics[width=0.6\textwidth]{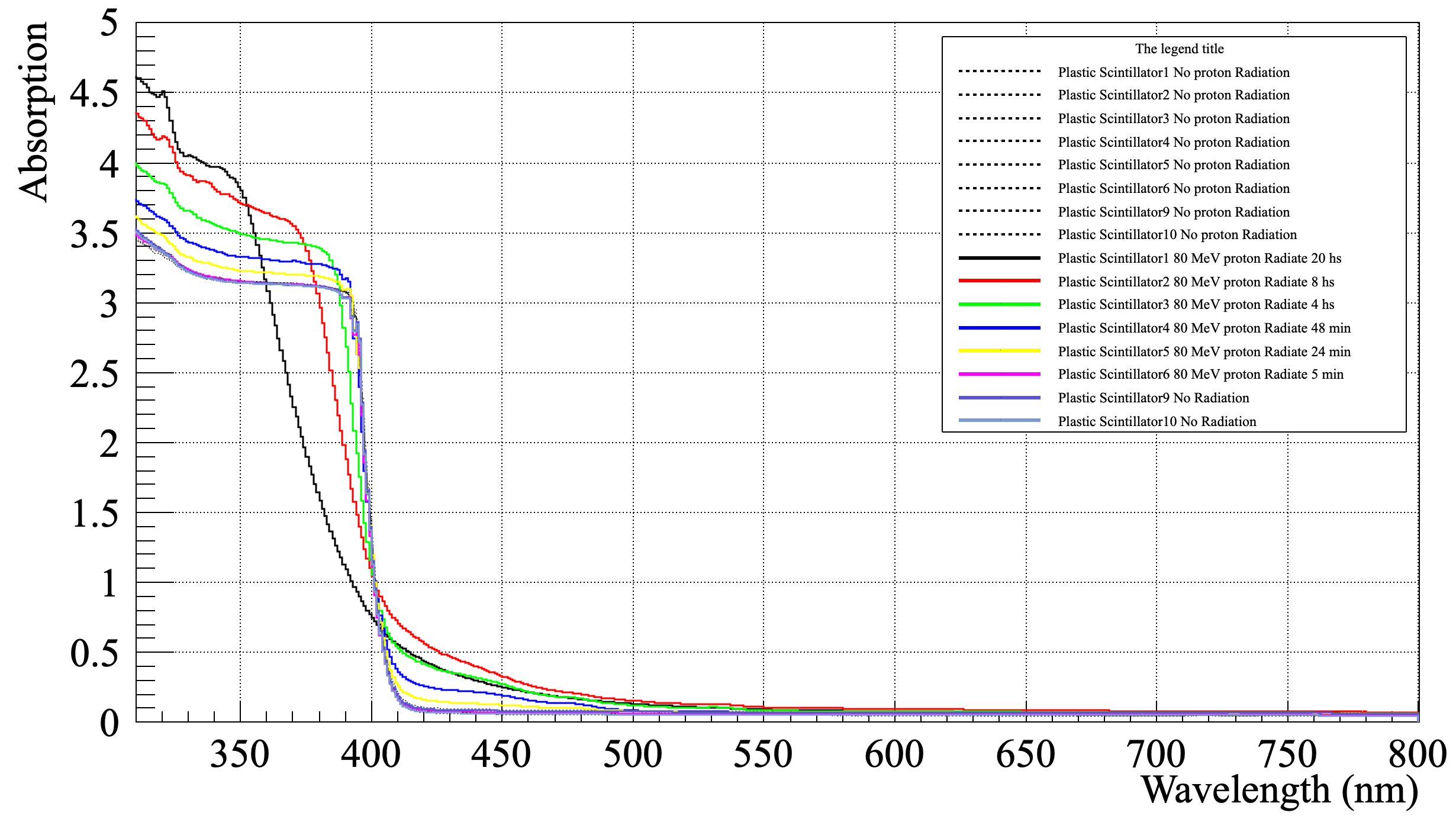}
  \caption{Comparison of absorption spectra of BC408 samples before and after proton irradiation.}
  \label{Fig3_Absor_Af}
\end{figure}

Figure \ref{Fig3_Absor_Af} shows the absorption spectra of eight BC408 samples after proton irradiation. Compared to the absorption spectra (black dotted line) of the pre-irradiated samples, the ultraviolet light from 300 nm to 350 nm experiences a greater absorption with an increase in irradiation time. Within the wavelength range of 350 nm to 400 nm, the starting wavelength value for a sharp change in absorption reduces with a longer irradiation time, indicating that the sample's capacity to absorb long-wave light decreases with longer irradiation time. In the visible light range above 400 nm, samples irradiated by protons have a higher light absorption ability, but the sample irradiated for 8 hours displaying greater absorbance than the one irradiated for 20 hours, which is related to the formation of a yellow core (figure \ref{Fig3_BC408_color_1}).

In the wavelength range of 300 nm to 800 nm, the absorbance of light in the near ultraviolet region (300 nm to 400 nm), visible blue, and blue-green spectrum has significantly increased after the irradiation. However, the absorption of green, yellow, and red light has slightly increased. Thus, the sample appears yellow when green, yellow, and red lights are overlaid.

\subsection{Fluorescence spectroscopy}

\begin{figure}[!htbp]
  \centering
  \includegraphics[width=0.6\textwidth]{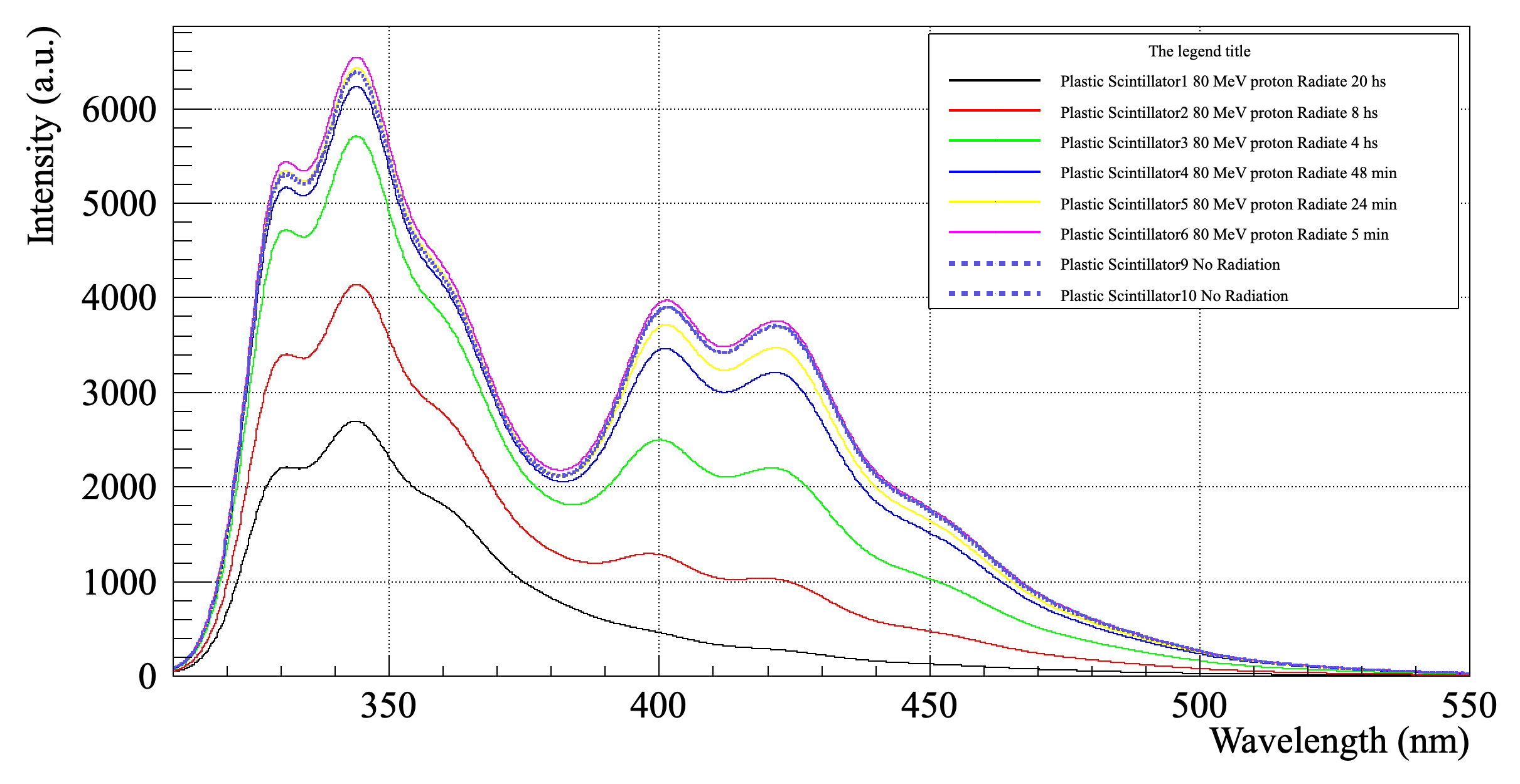}
  \caption{Comparison of fluorescence spectra of BC408 samples before and after proton irradiation.}
  \label{Fig3_Emis_BEAF}
\end{figure}

Figure \ref{Fig3_Emis_BEAF} shows the fluorescence spectra of eight BC408 samples of proton irradiated and unirradiated. 
The fluorescence spectra curves reveal several peak characteristics, with the peak position for different samples remaining relatively stable.
The Plastic9 and Plastic10 samples represent the control group and are identified by the purple dotted line, while the other solid lines are samples subjected to varying levels of irradiation.

Although the composition of BC408's primary and secondary flours remains a commercial secret, the fluorescence spectrum can be briefly analyzed based on the plastic scintillator's fluorescence mechanism. The fluorescence wavelength range from 310 nm to 380 nm corresponds with PVT bases and primary fluorescence, and the wavelength range from 380 nm to 500 nm is related to the secondary flour as a wavelength shifter.

As shown in the figure, the Plastic6 sample irradiated by $5.14\times10^3$ $\mathrm{Gy/cm^3}$ proton, is basically consistent with the control group, indicating minimal changes in fluorescence mechanism at low levels of absorbed dose. However, compared to the control group, the peak value of fluorescence spectra of Plastic1 with an absorbed dose of $1.23\times10^6$ $\mathrm{Gy/cm^3}$ decreased by 57.8\% and plastic2 with $4.94\times10^5$ $\mathrm{Gy/cm^3}$ decreased by 35.1\%. Consequently, when proton irradiates plastic scintillators, when the absorbed dose reaches a certain threshold, it will damage the composition of plastic scintillators and affect the plastic scintillators and the longer the irradiation time, the weaker the fluorescence intensity.

\subsection{Light yield testing in response to $^{241}\mathbf{Am}$}

\begin{figure}[!htbp]
  \centering
  \includegraphics[width=0.8\textwidth]{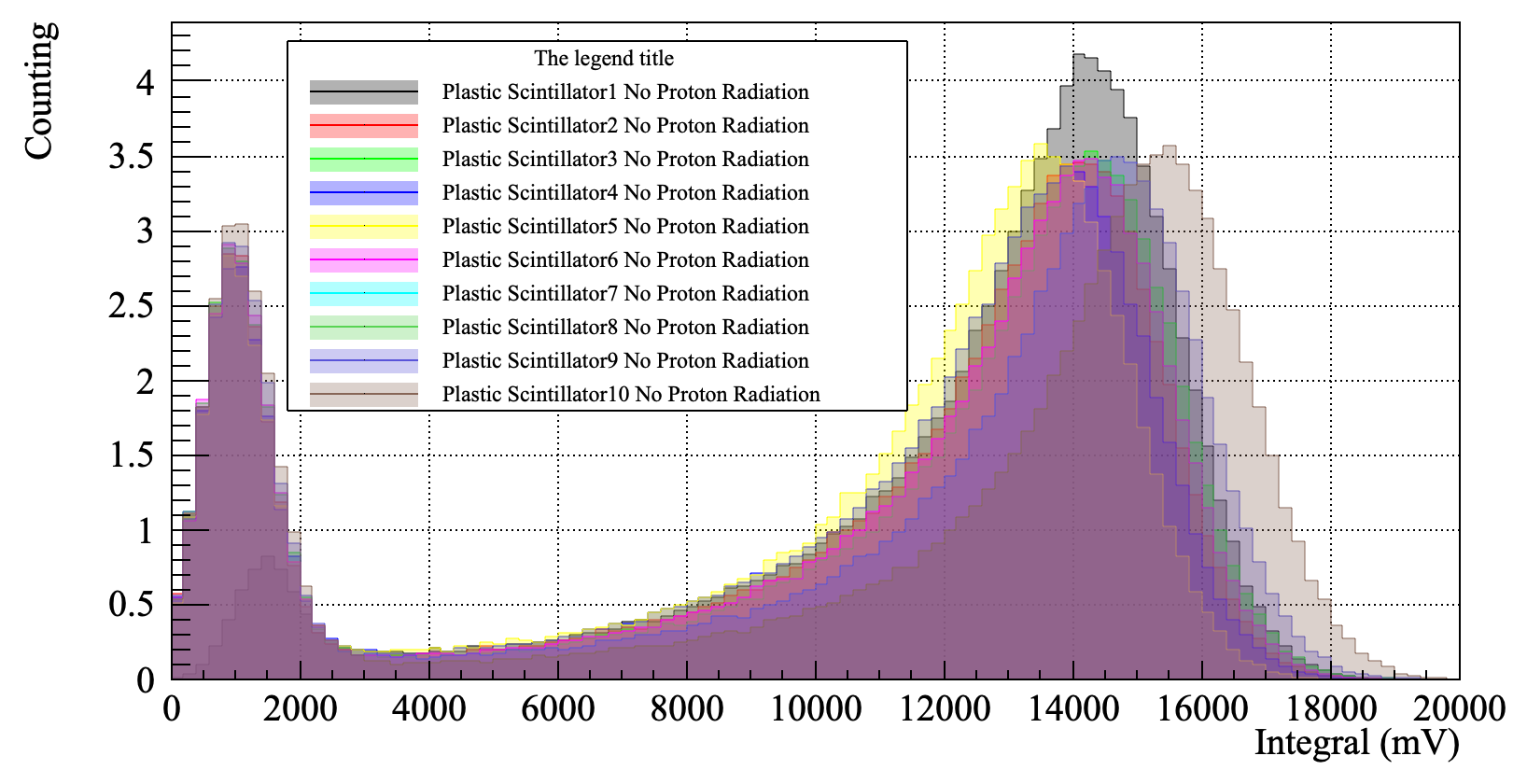}
  \caption{Unirradiated plastic scintillators responds to $^{241}\mathrm{Am} $radiation source.}
  \label{Fig3_Plastic_before_inter}
\end{figure}

Figure \ref{Fig3_Plastic_before_inter} shows the integral pulse spectra of 5.5 MeV $\alpha$ particle emitted from a $^{241}\mathrm{Am}$ source, measured using unirradiated BC408 samples.
The background peak is observed at a level of approximately 1000 mV. Table \ref{tab:plasic_da} lists the peak values of eight unirradiated samples, which exhibit a consistent response to 5.5 MeV alpha particles, with an average value of $14300\pm 595$ mV.

\begin{figure}[!htbp]
  \centering
  \includegraphics[width=0.8\textwidth]{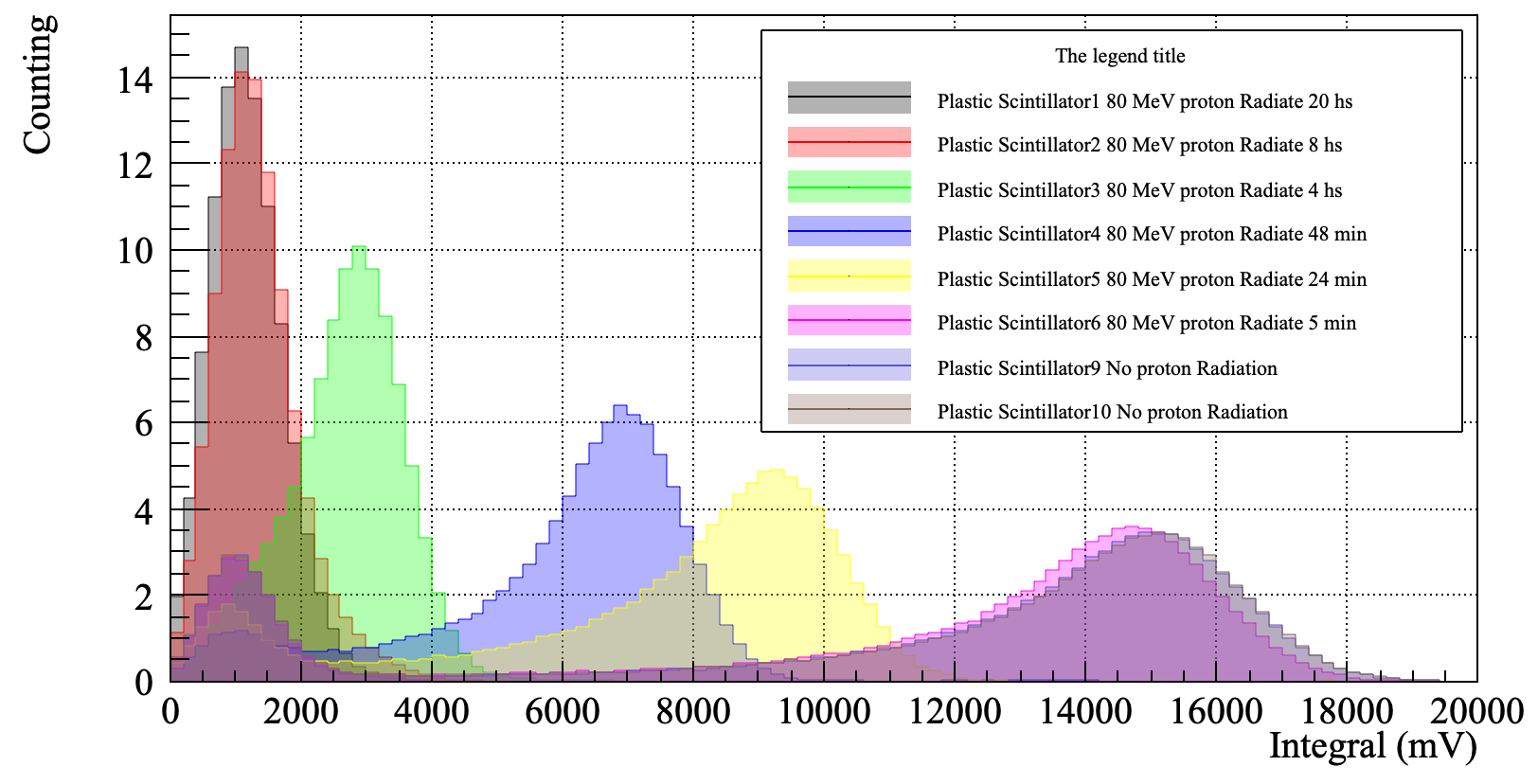}
  \caption{The response of plastic scintillators to $^{241}\mathrm{Am} $radiation source after proton irradiation.}
  \label{Fig3_Plastic_after_inter}
\end{figure}

Figure \ref{Fig3_Plastic_after_inter} displays the results obtained from irradiated samples, indicating a decline in performance with increasing proton irradiation. Notably, the longer the protons irradiate the BC408 samples, the weaker its response to $\alpha$ particles becomes, as evidenced by Table \ref{tab:plasic_da}. 
After absorbing a dose of $1.23\times10^6$ $\mathrm{Gy/cm^3}$, the scintillator sample Plastic1 exhibits a 92.2\% decrease in response to $\alpha$ particles compared to the control group, with its peak value detected at the background signal level, indicating its complete damaged. Similarly, sample Plastic2 loses its ability to detect $\alpha$ particles after absorbing $4.94\times10^5$ $\mathrm{Gy/cm^3}$ of protons.
The data in Table \ref{tab3:PlasticSetup} suggests that BC408 scintillators degrade completely after being irradiated with 1.6 GeV high-energy protons at a flux of approximately $1.56\times10^{15} \thinspace \mathrm{p/cm^2}$. Plastic3, Plastic4 and Plastic5 exhibit a 79.7\%, 50.4\% and 31.1\% decrease in response to $\alpha$ particles after absorbing dose levels of $2.47\times10^5$ $\mathrm{Gy/cm^3}$, $4.94\times10^4$ $\mathrm{Gy/cm^3}$ and $2.47\times10^4$ $\mathrm{Gy/cm^3}$, respectively. However, Plastic6, which absorbed a dose of $5.14\times10^3$ $\mathrm{Gy/cm^3}$, exhibited no significant response change before or after irradiation, consistent with the control group response. In conclusion, the absorption and fluorescence spectra of Plastic6 show no significant changes before and after irradiation, demonstrating that the radiation threshold of BC408 scintillators likely exceeds $5.14\times10^3$ $\mathrm{Gy/cm^3}$. Combining these results with the data presented in Table \ref{tab3:PlasticSetup} suggests that BC408 scintillators can withstand a flux of 1.6 GeV high-energy protons on the order of $1.63\times10^{13} \thinspace \mathrm{p/cm^2}$, while maintaining their properties.

\begin{table}[!htbp]
\centering
\caption{Response of BC408 scintillators to $^{241}\mathrm{Am}$ radiation source after proton irradiation.}
\label{tab:plasic_da}
\begin{tabular}{l|l|l|l|l|l}
\toprule
\multirow{1}{*}{Condition} & not irradiated                                        & \multicolumn{4}{c}{80 MeV proton irritated} \\ \hline
Plastic ID & \makecell{Peak of 5.5 MeV \\ $\alpha$ particle (mV) } & Irradiated Time &  \makecell{Absorbed dose\\($\mathrm{Gy/cm^3}$)} &  
          \makecell{Peak of 5.5 MeV \\ $\alpha$ particle (mV)} & \makecell{Peak change \\ (\%)} \\ \midrule
Plastic1  & 14100 & 20 hours & $1.23\times10^6$ & 1100 & 92.2 \\
Plastic2  & 14100 & 8 hours  & $4.94\times10^5$ & 1100 & 92.2 \\
Plastic3  & 14300 & 4 hours  & $2.47\times10^5$ & 2900 & 79.7 \\
Plastic4  & 13900 & 48 mins  & $4.94\times10^4$  & 6900 & 50.4 \\
Plastic5  & 13500 & 24 mins  & $2.47\times10^4$  & 9300 & 31.1 \\
Plastic6  & 14300 & 5 mins   & $5.14\times10^3$   & 14700 & -2.8 \\
Plastic9  & 14700 & not irradiated & not irradiated & 14900 & -1.4 \\
Plastic10 & 15500 & not irradiated & not irradiated & 15100 & 2.6 \\
\bottomrule
\end{tabular}
\end{table}

\section{Conclusion}

In this study, the proton irradiation damage effect of BC408 plastic scintillator was studied. 
Absorption, fluorescence spectroscopy and light yield tests demonstrated that BC408 has a certain resistance to proton irradiation and maintains its performance when the absorbed dose is below $5.14\times10^3$ $\mathrm{Gy/cm^3}$. When the absorbed dose is more than $2.47\times10^4$ $\mathrm{Gy/cm^3}$, the performance of BC408 begins to decrease significantly. By analyzing the changes of absorption and fluorescence spectrum, it is shown that proton irradiation will damage the luminous composition of plastic scintillator and change its luminous mechanism. When the absorbed dose is greater than $4.94\times10^5$ $\mathrm{Gy/cm^3}$, the BC408 has lost ability to response the $\alpha$ particle, and its performance damage completely. However, once the absorbed dose exceeds $2.47\times10^4$ $\mathrm{Gy/cm^3}$, BC408's performance deteriorates significantly. Analysis of the absorption and fluorescence spectrum changes reveals that proton irradiation damages the luminous composition of the plastic scintillator and alters its luminous mechanism. Specifically, if the absorbed dose surpasses $4.94\times10^5$ $\mathrm{Gy/cm^3}$, BC408 loses its ability to respond to $\alpha$ particles, causing complete performance degradation. When using BC408 as a flux beam detector on the CSNS-II-Up 1.6 GeV high-energy proton beam experiment platform, it remains stable as long as the absorbed dose is less than $5.14\times10^3$ $\mathrm{Gy/cm^3}$, which equates to absorbing a maximum of $1.63\times 10^{13}\,\mathrm{p/cm^2}$ 1.6 GeV protons. Finally, according to the current high-energy proton beam line design parameters, the flux of 1.6 GeV protons is about $10^8 \thinspace \mathrm{p/cm^2}$, the BC408 beam detector could potentially be used as the 1.6 GeV high-energy proton beam detector in the CSNS Phase-II upgrade project.

\section{Acknowledgments}
This work was supported by the National Natural Science Foundation of China (No. U2032165).

\section*{References}


\end{document}